# Spin-transfer switching and low-field precession in exchange-biased spin valve nano-pillars

M. C. Wu, A. Aziz, D. Morecroft, M. G. Blamire

Department of Materials Science, University of Cambridge,

Pembroke Street, Cambridge, CB2 3QZ, United Kingdom

M. C. Hickey, M. Ali, G. Burnell, B.J. Hickey

School of Physics and Astronomy, University of Leeds, Leeds, LS2

9JT, United Kingdom

Using a three-dimensional focused-ion beam lithography process we have fabricated nanopillar devices which show spin transfer torque switching at zero external magnetic fields. Under a small in-plane external bias field, a field-dependent peak in the differential resistance versus current is observed similar to that reported in asymmetrical nanopillar devices. This is interpreted as evidence for the low-field excitation of spin waves which in our case is attributed to a spin-scattering asymmetry enhanced by the IrMn exchange bias layer coupled to a relatively thin CoFe fixed layer.

In 1996, Slonczewski and Berger predicted that a spin-polarized current, which is caused to flow perpendicular to plane between a relatively thick ferromagnetic layer through a non-magnetic layer to a thin free nanomagnetic layer, could transfer spin momentum from the current to the free layer.[1,2] Depending on the direction of the spin current flow, the spin transfer effect can either force the free layer into parallel (P) or antiparallel (AP) alignment compared with the fixed layer when the spin transfer torque (STT) is strong enough to overcome the coercive field of the free layer. Furthermore, when an external magnetic field is applied, the effect of spin transfer can be the excitation of spin-wave precessional modes in the free layer.[3-6] There have been numerous studies which demonstrate the basic principle of STT switching and spin wave excitation.[5-15] Until recently, spin-wave excitation required the application of large fields (~1 T) to destabilise the AP state of the device, but recently zero-field spin-wave excitation has been reported in devices in which the spin-transfer torque can destabilise both the parallel and antiparallel alignment of the ferromagnetic layers.[16] This Letter reports similar behavior in devices for which the necessary spin-scattering asymmetry is enhanced by the use of an exchange bias layer coupled to a thin fixed layer rather than by making the free and fixed layers from different materials.

STT has been observed in a number of device geometries, which include mechanical point contacts,[3,6,10,11] lithographically-defined point contacts,[4] and lithographically-defined nanopillars.[5,7,8,12-15] The devices reported here have been fabricated by a simple and reliable procedure based on using 3-D focused ion beam (FIB) milling.[17] Since in this process all the metal layers are deposited in a single ultra high vacuum cycle, excellent interface cleanliness is achieved. The technique provides a very reliable one stage process for fabricating nano-pillars for investigating the spin transfer torque effects in magnetic thin films.

Devices have been fabricated using the following steps. Firstly, thin film heterostructures were deposited onto thermally oxidized Si substrate in an ultrahigh vacuum sputtering system with the base pressure below ~$5\times10^{-8}$ mbar. The structure of the multilayered thin film is Ta(5)/Cu(200)/CoFe(3)/Cu(6)/CoFe(6)/IrMn(10)/Cu(200)/Ta(5) (thickness in nm). The magnetic properties of the deposited thin film were characterized using vibrating sample magnetometery and showed a significant exchange bias of ~10 mT. Secondly, using conventional photolithography and Ar ion milling, the multilayer thin film was patterned into a 4 μm wide track with appropriate current and voltage leads. All further processing was performed using FIB milling on a custom-built 45° wedge holder, firstly to narrow the optically-defined tracks down to 150 nm and then, to achieve the required current perpendicular to plane (CPP) geometry, lateral slots were milled as shown in Fig. 1(a). A micrograph of a completed device is shown in Fig. 1(b).

Transport measurements were performed using a four terminal ac lock-in technique. The dynamic resistance (dV/dI) of the nano-pillar was measured using an ac current excitation of 200 μA rms at 77 Hz. A dc bias current was simultaneously applied during the dV/dI measurement, with the positive (negative) direction corresponding to electron flowing from the fixed (free) to the free (fixed) layer. The dynamic resistance was measured as a function of the magnetic field and dc bias current, with the in-plane magnetic field applied along the geometric easy axis (long axis of the rectangle). Fig. 2(a) shows a magnetoresistance (MR) loop with zero bias current for an exchange biased spin valve nano-pillar with a size of 150 nm by 200 nm. At low fields, a high resistance state is generally observed, and a field of ~20 mT is required to align the magnetization of the layers and the resistance is a minimum.

Micromagnetic simulations of this device were performed using the three-dimensional Object Oriented Micromagnetic Framework (OOMMF) software.[18] The nano-pillar was divided into $5 \times 5 \times 3$ nm$^3$ cells. The saturation magnetisation and uniaxial anisotropy CoFe were set to $1.3 \times 10^6$ Am$^{-1}$ and $5 \times 10^4$ J m$^{-3}$ respectively, and the damping coefficient was set to 0.5 to obtain rapid convergence.[19,20] The inter-layer exchange energy was set to zero and the biasing field was set to 10 mT at the interface between IrMn and CoFe. The relative angle between the magnetic cells in the hard and soft layer was derived from the micromagnetic simulations, and the MR was assigned as 0 or 1 when the layers were parallel or antiparallel respectively. The simulated MR result (Fig. 2(b)) shows good agreement with the experimental result (Fig. 2(a)).

Our devices show clear STT switching in zero external magnetic field at room temperature, as shown in Fig 3. For the device shown, the switch to the P state occurs at $I_{AP-P}$= 10.1 mA and back to antiparallel state at $I_{P-AP}$= -1.2 mA. The corresponding change in resistance at zero bias dc current is similar to the change in resistance at zero external applied field in Fig. 2(a), confirming that the magnetization is fully reversed by the spin-polarized current injection. The critical current density is estimated to be $3.4 \times 10^7$ A/cm$^2$ (for $I_{AP-P}$) and $-0.4 \times 10^7$ A/cm$^2$ (for $I_{P-AP}$). The current polarities follow the Slonczewski model but $|I_{AP-P}| > |I_{P-AP}|$. This is apparently in contrast with the Slonczewski model, which shows that $|I_{AP-P}| < |I_{P-AP}|$. However, this difference can be explained by the presence of a large magnetostatic coupling which favours AP state.

Despite the use of a relatively thin fixed layer, the shape and structure of our nano-pillar devices results in a strong magnetostatic coupling between the ferromagnetic layers making the AP state more energetically favorable at zero field. This would tend to decrease $I_{P-AP}$ and increase $I_{AP-P}$ and so a reversed switching anisotropy is possible for sufficiently large magnetostatic coupling. To see if the magnitude of this field is reasonable to explain this effect, we can relate the effective field associated with the difference in $I_{P-AP}$ and $I_{AP-P}$ to the difference in the applied fields required to switch the free layer. In Fig. 2(a) we plot the minor loop resulting from the switching of the free layer between the P and AP states; the switching hysteresis, i.e. twice the coercive field, is ~10 mT. The midpoint of this loop (~12 mT) represents the magnetostatic field from the fixed layer. Thus the minor loop offset is equal to its width which is at least qualitatively similar to the STT switching behavior. We can rule out magnetic (Oersted field) switching because we calculate that the field generated by a 10 mA drive current is less than 8 mT, which is not sufficient to overcome the switching field of the free layer (~20 mT). Therefore we can be confident that the switching observed in our experiments is induced by STT but in the presence of a large magnetostatic coupling which makes the AP state energetically favorable.

When a sufficiently large external bias field is applied so that the nano-pillar is in a parallel configuration, the irreversible switching is replaced by a peak in the differential resistance whose current depends on the applied field (Fig. 4). We interpret this peak as the critical excitation current for spin-waves; the peak current ($I_p$) is linearly dependent on field as shown in the inset to Fig. 4. For comparison we also show the field-dependence of similar peaks extracted from Boulle *et al.* which were shown directly to correspond to spin-wave excitation currents.[16] The similarity of the $I_c(H)$ behavior, the field range in which the peaks occur, and the fact that the peaks occur for positive currents which tend to stabilise the parallel configuration for conventional STT devices strongly suggests a similar underlying origin which is different from the more conventional high field spin-wave regime modelled by Slonzewski[21] and experimentally observed by a number of groups.[3,10]

Spin-wave excitation at low field seems to require a so-called "wavy" angular dependence of the torque on the misorientation angle between the moments of the two layers which can arise because of spin accumulation in the normal metal spacer;[22] this can be obtained by an asymmetry of the spin diffusion length in the two ferromagnetic layers (Co and Py).[16] Our devices are symmetric to the extent that both layers are CoFe, but the asymmetry is enhanced by the presence of the IrMn exchange bias layer coupled to a relatively thin fixed layer. The spin-diffusion length of CoFe is 12 nm[23] and so both FM layers are substantially thinner than this; however, the strong spin-scattering at the IrMn/CoFe interface effectively reduces the spin-diffusion length in the fixed layer and enhances the necessary asymmetry.

The peak structures in our device occur at low, but not zero-field; this is presumably due to the strong magnetostatic coupling in our pillars so that finite field is required to obtain a parallel state. The extrapolation of the linear fit to $I_p(H)$ to zero current in fact corresponds to the coupling field of around 15 mT so that in the absence of this coupling field (for example through the use of a synthetic antiferromagnetic fixed layer) our data and that of Boulle et al. would be essentially identical (see inset to Fig. 4).

In summary, a 3-D FIB process has been used to investigate the spin transfer effect in exchange-biased spin valve nano-pillars. This technology provides an efficient method to fabricate reliable magnetic spin transfer devices. We have observed low-field peak structures consistent with spin wave excitation, and we believe that the use of spin-scattering interfaces may provide a more flexible method of enhancing the asymmetry necessary for zero field spin-wave excitation for microwave radiation applications.

This work was supported by U.K. Engineering and Physical Sciences Research Council under the Spin@RT consortium and M. C. Wu would like to acknowledge the financial support of the Ministry of Education of Taiwan and a useful discussion on the exchange-biased thin film from Rantej Bali.

References:

1   J.C. Slonczewski, J. Magn. Magn. Mater. 159, L1 (1996).

2   L. Berger, Phys. Rev. B 54, 9353 (1996).

3   M. Tsoi, A. G. M. Jansen, J. Bass, W.-C. Chiang, M. Seck, V. Tsoi, and P. Wyder, Phys. Rev. Lett. 80, 4281 (1998).

4   E. B. Myers, D. C. Ralph, J. A. Katine, R. N. Louie, and R. A. Buhrman, Science 285, 867 (1999).

5   J. A. Katine, F. J. Albert, R. A. Buhrman, E. B. Myers, and D. C. Ralph, Phys. Rev. Lett. 84, 3149 (2000).

6   S. M. Rezende, F. M. de Aguiar, M. A. Lucena, and A. Azevedo, Phys. Rev. Lett. 84, 4212 (2000).

7   F. J. Albert, J. A. Katine, R. A. Buhrman, and D. C. Ralph, Appl. Phys. Lett. 77, 3809 (2000).

8   S. Urazhdin, N. O. Birge, W. P. Pratt, and J. Bass, Phys. Rev. Lett. 91, 146803 (2003).

9   J. Akerman, Science 308, 508 (2005).

10  Y. Ji, C. L. Chien, and M. D. Stiles, Phys. Rev. Lett. 90, 106601 (2003).

11  W. H. Rippard, M. R. Pufall, and T. J. Silva, Appl. Phys. Lett. 82, 1260 (2003).

12  M. Tsoi, J. Z. Sun, and S. S. P. Parkin, Phys. Rev. Lett. 93, 036602 (2004).

13  M. AlHajDarwish, H. Kurt, S. Urazhdin, A. Fert, R. Loloee, W. P. Pratt, and J. Bass, Phys. Rev. Lett. 93, 157203 (2004).

14  Y. Jiang, S. Abe, T. Ochiai, T. Nozaki, A. Hirohata, N. Tezuka, and K. Inomata, Phys. Rev. Lett. 92, 167204 (2004).

15  X. Jiang, L. Gao, J. Z. Sun, and S. S. P. Parkin, Phys. Rev. Lett. 97, 217202 (2006).

16  O. Boulle, V. Cros, J. Grollier, L. G. Pereira, C. Deranlot, F. Petroff, G. Faini, J. Barnas, and A. Fert, Nature Physics 3, 492 (2007).

17  C. Bell, G. Burnell, D. J. Kang, R. H. Hadfield, M. J. Kappers and M. G. Blamire, Nanotechnology 14, 630 (2003).

18  http://math.nist.gov/oommf. Version 1.2.0.4 of the software was used.

19  S. H. Lim, S. H. Han, K. H. Shin, and H. J. Kim, J. Magn. Magn. Mater. 223, 192 (2001).

20  S. Bedanta, E. Kentzinger, O. Petracic, W. Kleemann, U. Rücker, Th. Brückel, A. Paul, S. Cardoso, and P. P. Freitas, Phys. Rev. B 74, 054426 (2006).

21  J.C. Slonczewski, J. Magn. Magn. Mater. 195, L261 (1999).

22  M. Gmitra and J. Barnas, Appl. Phys. Lett. 89, 223121 (2006).

23  J. Bass and W. P. Pratt, J Phys: Condens. Matter 19, 183201 (2007).

Figure captions:

Fig. 1 (Color online) (a) Scheme of the device fabrication by 3-D focused ion beam milling and (b) FIB micrograph of the nano-pillar and diagram of the current flow direction.

Fig. 2 (Color online) (a) Experimental magnetoresistance of the nano-pillar measured at room temperature, the dashed line shows a minor loop from +60 mT → 0 → +60 mT. (b) Simulated magnetoresistance of the nano-pillar.

Fig. 3 (Color online) Dynamic resistance of the nano-pillar as a function of DC current measured with no external field and at room temperature.

Fig. 4 (Color online) Measurements of the dynamic resistance of the nano-pillar as a function of DC bias current, for different applied magnetic fields. Measurements were taken at room temperature. The measurements all fall on to the same thermal background which was subtracted. Inset shows the peak current density (in $Acm^{-2}$) vs field for our device (blue triangles), and data from Boulle et al (see Ref. 16) (purple squares).

Figure 1:

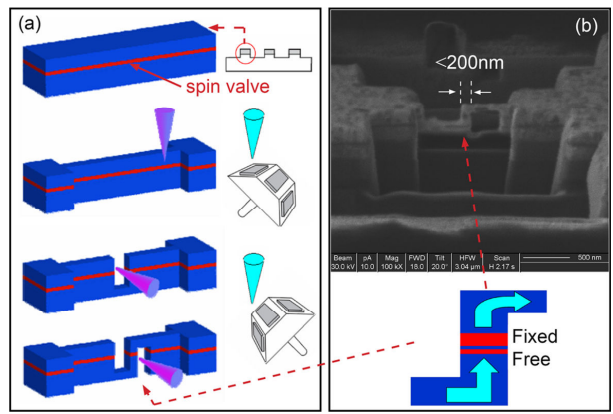

Figure 2:

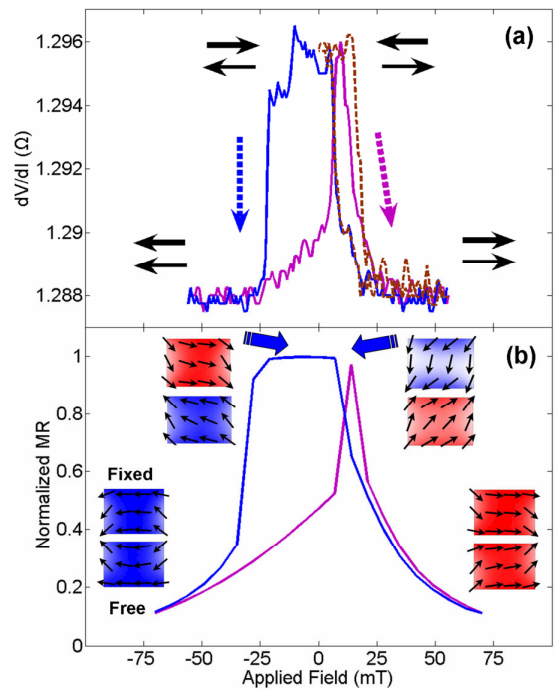

Figure 3:

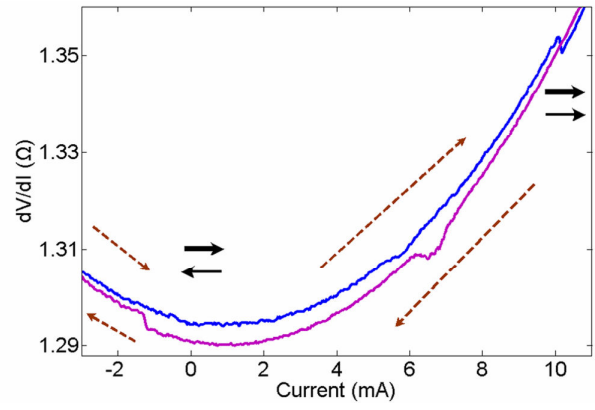

Figure 4:

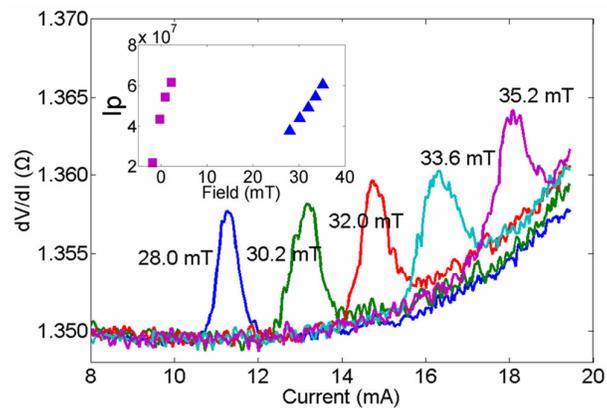